\documentclass[10pt,aps,twocolumn,prl,amsmath,floatfix,longbibliography,footinbib,bibnotes,a4paper]{revtex4-2}
\pdfoutput=1        
\usepackage{graphicx}
\usepackage{microtype}
\usepackage{hyperref}
\def\etal{{\itshape et al. \/}}

\def\rmd{{\rm d}}

\def\DOS{\mathop{\mathrm{DOS}}\nolimits}

\begin{document}

\title{The magnetic exciton of EuS revealed by resonant inelastic
  x-ray scattering}

\author{Lucia Amidani}
\thanks{Correspondence and requests for materials should be addressed to
L.A. or J.K.}
\email{lucia.amidani@esrf.fr}
\affiliation{The Rossendorf Beamline (ROBL) at the ESRF, 71 Avenue des
  Martyrs, Grenoble 38043, France}
\affiliation{Institute of Resource Ecology, Helmholtz-Zentrum
  Dresden-Rossendorf (HZDR), Bautzner Landstra\ss e 400, 01328
  Dresden, Germany}
\author{Jonas J. Joos}
\affiliation{LumiLab, Department of Solid State Sciences, Ghent
  University, Krijgslaan 281-S1, B-9000 Gent, Belgium}
\author{Pieter Glatzel}
\affiliation{ESRF -- The European Synchrotron, 71 Avenue des Martyrs,
  38000 Grenoble, France}
\author{Jind\v{r}ich  Koloren{\v{c}}}
\email{kolorenc@fzu.cz}
\affiliation{Institute of Physics (FZU), Czech Academy of Sciences, Na
  Slovance 2, 182 00 Prague, Czech Republic}

\hypersetup{
  pdfauthor={Lucia Amidani (https://orcid.org/0000-0003-2234-4173),
    Jonas Joos (https://orcid.org/0000-0002-7869-2217),
    Pieter Glatzel (https://orcid.org/0000-0001-6532-8144),
    Jindrich Kolorenc (https://orcid.org/0000-0003-2627-8302)},
  pdftitle={The magnetic exciton of EuS revealed by resonant inelastic
    x-ray scattering},
  pdfkeywords={europium, sulfide, EuS, exciton, RIXS, optical
    absorption, DFT, DMFT}
}

\date{October 27, 2023}

\begin{abstract}
We report the valence-to-core resonant inelastic x-ray scattering
(RIXS) of EuS measured at the L${}_3$ edge
of Eu. The obtained data reveal two sets of excitations: one set is composed
of a hole in the S~3p bands and an electron excited to extended Eu 5d
band states, the other is made up from a hole in the Eu 4f states and an
electron in localized Eu 5d states bound to the 4f hole by its
Coulomb potential. The delocalized
excitations arise from the dipole-allowed 5d${}\to{}$2p emissions,
whereas the localized excitations result from the dipole-forbidden
(quadrupole-allowed) 4f${}\to{}$2p emissions. Both these emission channels have
a comparable intensity thanks to a small number of occupied 5d states
($\approx 0.6$) combined with a large number of occupied 4f states
(seven). We identify the localized electron-hole pairs with the
``magnetic excitons'' suggested in the past as an interpretation of the
sharp features seen in the optical absorption spectra. Our
observations provide a direct experimental evidence of these excitons
which has been missing up to now.
\end{abstract}

\maketitle

%
%

Europium sulfide (EuS) belongs to the europium monochalcogenides
series (EuX, X = O, S, Se, Te), which represents a rare example of
intrinsic magnetic semiconductors \cite{mauger1986}. The coupling of
semiconducting and magnetic properties in the members of the EuX
series is attractive for spintronics and magneto-optics
applications. EuX crystallize in the rock salt structure, with the
Eu$^{2+}$ ions having their 4f shell half-filled (4f${}^7$ configuration) and carrying
purely spin magnetic moment, which makes EuX prototypical examples of
Heisenberg magnets. The semiconducting gap is found between the localized
and occupied 4f states and the empty conduction band of predominantly Eu 5d
character, while the S 3p states constitute the occupied valence band
located below the occupied 4f states \cite{mauger1986}.

The discovery of the Eu monochalcogenides in the 1960s was accompanied by a strong scientific
interest for their potential use in spin-related technologies
\cite{matthias1961}. When the impossibility to raise the
Curie temperature up to room temperature by doping became clear, the interest in
EuX turned toward more fundamental investigations of these model
systems. In the last two decades, however, the interest in EuX was
renewed by the discovery of new properties, like those of EuX
nanoparticles intended for magneto-optical devices \cite{huxter2008,
  hasegawa2013, hasegawa2015, asuigui2020}, the interface magnetism
induced by coupling EuX with a topological insulator
\cite{katmis2016}, the possibility to raise the Curie temperature in
strained multilayered structures \cite{lechner2005, ingle2008}, or
the demonstration that optical control can be used to induce EuX
magnetization on the ultrafast timescale \cite{matsubara2015,
  matsubara2019, kats2023}.

The renewed interest in EuX stimulated novel fundamental
investigations, focusing in particular on the study of the electronic
structure across the Curie temperature to understand the exchange
mechanism responsible for the ferromagnetic ordering
\cite{souza-neto2009, miyazaki2009, riley2018, fedorov2021}. Less
attention has been paid to an unsettled debate about the interpretation
of the EuX optical absorption. The absorption
spectrum of EuX is characterized by two peaks, the first being located
at the onset
and the second on the rising edge of the spectrum
\cite{guntherodt1971}.
While there seems to be a general consensus that these two
peaks originate from transitions, in which a 4f valence electron is
excited to the crystal-field-split 5d($t_{2g}$) and 5d($e_g$)
states, the spatial extent of these excited states is debated.

The
interpretation proposed by Wachter and collaborators assumes that the
excited electron is itinerant and resides in a \emph{delocalized}
single-particle Bloch state. The
shape of the absorption spectrum then implies that the 5d bands in
EuS are very narrow, with their width being smaller than the crystal-field
splitting \cite{guntherodt1971}. This picture was criticized by Kasuya
and collaborators who argue that the Coulomb attraction of the hole
created in the 4f shell prevents the excited electron from leaving the
atom. Instead, \emph{localized} many-body states
4f${}^6$5d${}^1$($t_{2g})$ and 4f${}^6$5d${}^1$($e_g)$, termed
``magnetic excitons'', are formed \cite{kasuya1968, kasuya1972}. Such
excitons induce sharp features in the optical absorption spectrum
without requiring all 5d bands to be narrow.

In the absence of a direct experimental proof, the debate about the nature of
the absorption spectrum of EuX was never satisfactorily resolved
\cite{kasuya1999,
wachter2013}. In most works and reviews dealing with EuX, the
interpretation of Watcher \etal \cite{wachter1972} is given as
correct \cite{mauger1986, huxter2008}. Only few Japanese groups,
investigating the magneto-optical properties of EuX nanoparticles
\cite{hasegawa2013, hasegawa2015} and the ultrafast magnetization of
EuX through laser control \cite{matsubara2019}, still consider
the Kasuya's model as correct and interpret their data accordingly.

%
%

\begin{figure}
\includegraphics[width=\linewidth]{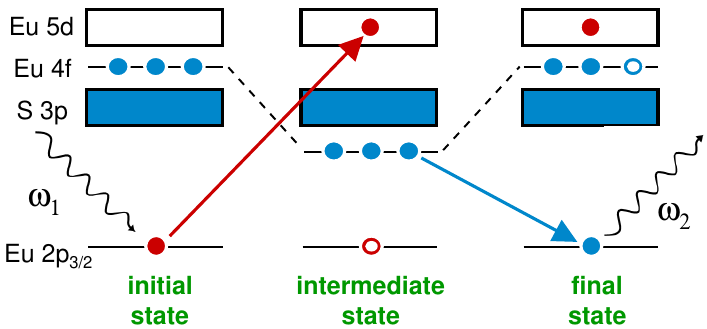}
\caption{\label{fig:RIXStransitions}Scheme of the RIXS process. The
  incoming x-ray $\omega_1$ excites a core electron to an unoccupied
  state. Then, a
  valence electron decays and fills the core hole, releasing the
  excess energy by emitting $\omega_2$. As a result, a valence
  electron is promoted to an unoccupied state.}
\end{figure}

In this Letter, we report the first experimental proof of the
excitonic nature of the onset of the absorption spectrum in
EuS. Experimental valence-to-core resonant inelastic x-ray scattering
(RIXS) at the Eu L${}_3$ edge, interpreted with the density functional
theory (DFT) and the dynamical mean-field theory (DMFT), demonstrates that
the two peaks of the optical absorption
spectrum are the localized 4f--5d excitons. In RIXS at the L${}_3$ edge, an Eu 2p${}_{3/2}$
core electron is excited to the unoccupied density of states of
the d~symmetry, and the intensity of the x-rays scattered in the
0~eV to 15~eV range of energy transfer is subsequently
detected. Figure~\ref{fig:RIXStransitions} shows the scheme of the RIXS 
process relevant to our case. The absorption of the x-ray photon
$\omega_1$ brings the system into an intermediate state with a
2p${}_{3/2}$ core hole that perturbs the valence electronic structure. Electrons
from occupied states in the valence-band region then fill the core hole and
the excess energy is released by an emission of a second x-ray
photon $\omega_2$. We consider dipole allowed as well as dipole
forbidden (but quadrupole allowed) emission channels in the following
discussion. The result of the RIXS process is a final state with an
electron transferred from the valence to the conduction band, analogous
to what is obtained in optical spectroscopy through the absorption of
a single photon in the UV--visible range.

\begin{figure*}
\includegraphics[width=\linewidth]{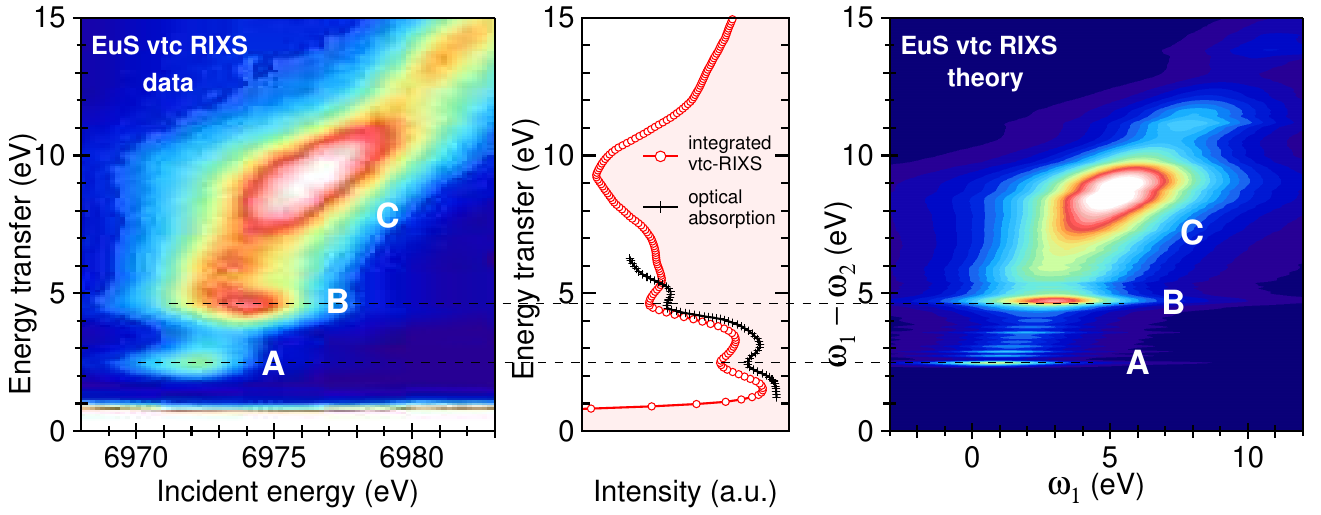}
\caption{\label{fig:RIXSmaps}a) experimental RIXS, b) comparison
  of the optical absorption spectrum of EuS from
  \cite{poulopoulos2012} with the integrated RIXS along the
  incident-energy axis, c) calculated RIXS including the dipole
  (Eu 2p${}\to{}$Eu 5d, Eu 5d${}\to{}$Eu 2p) and quadrupole (Eu
  2p${}\to{}$Eu 5d, Eu 4f${}\to{}$Eu 2p) emission. The dipole
  contribution reflects the extended 5d band states, the quadrupole
  contribution reveals the localized excitons.}
\end{figure*}

Figure~\ref{fig:RIXSmaps}a shows the experimental RIXS of EuS
collected on the ID26 beamline at the ESRF \cite{glatzel2021}. The
RIXS map reports the intensity of the $\omega_2$ scattered x-rays as a
function of the energy of the $\omega_1$ incident x-rays and of the
energy transferred to the system, that is, $\omega_1 - \omega_2$. Three
main features appear at positive energy transfer. Two sharp and
localized features at 2.5~eV and 4.5~eV energy transfer (labeled~A
and~B), and a broad and extended feature centered at 9~eV energy
transfer (labeled~C). By integrating the RIXS along the incident-energy
axis, we are summing over all the intermediate states reached by the
absorption of the $\omega_1$ x-ray photon (see
Fig.~\ref{fig:RIXStransitions}). The resulting curve corresponds to
all the final states reached by the RIXS process and can be compared
to the optical absorption. Figure~\ref{fig:RIXSmaps}b shows
the comparison between our integrated RIXS and the optical
absorption spectrum of EuS from \cite{poulopoulos2012}, to which a
0.1~eV shift has been applied to align it with our data.
The shapes of the two curves are indeed very close and the two peaks at
the onset of the optical absorption spectrum, the nature of which is
debated in literature, correspond to features~A and~B of our
RIXS. The optical spectrum stops before feature~C and hence a
direct comparison cannot be made. The
possibility to observe the peaks~A and~B in two dimensions with RIXS reveals
important details about their nature. It shows that they are
well-separated from feature~C and they are reachable only for a limited range of
incident energies at the onset of the Eu
L$_3$-edge x-ray absorption spectrum.

%
%

To understand the observed RIXS features, we modeled the
RIXS process with the simplified formula proposed by
Jim\'enez-Mier \etal \cite{jimenez1999} and further developed and
benchmarked against experimental data by Smolentsev
\etal \cite{smolentsev2011}. When the absorption of $\omega_1$ and
emission of $\omega_2$ can be disentangled, the direct RIXS process
can be described as an absorption followed by an emission and the
Kramers--Heisenberg formula \cite{kramers1925,sakurai_AQM_KHformula}
giving the RIXS intensity reduces to the convolution of the
unoccupied and occupied densities of states (DOS) projected to the
symmetry allowed by the electric dipole selection rules,
\begin{equation}
\label{eq_RIXS_dip}
I_D \propto \int \rmd\epsilon\,
\frac{\DOS_{\text{5d}}^\text{occ}(\epsilon)\,
\DOS_{\text{5d}}^{\text{empty}}(\epsilon+\omega_1-\omega_2)}%
{(\epsilon-\omega _2-\epsilon_\text{2p})^2+\Gamma_\text{2p}^2/4}\,.
\end{equation}
Here $\omega_1$ and $\omega_2$ are the energies of the absorbed and
emitted photons, $\epsilon_\text{2p}$ is the energy of the 2p${}_{2/3}$
core level, and $\Gamma_\text{2p}=3.91$~eV is its width due to lifetime
broadening (the full width at half maximum, value adopted from
\cite{krause1979}).

We calculated the EuS band structure and the DOSes to be inserted into
Eq.~\eqref{eq_RIXS_dip} with the DFT+DMFT method, which solves a
multi-band Hubbard model built on top of the DFT band structure. The
Coulomb repulsion added to the Eu 4f shells is parametrized by four
Slater parameters $F_k$, the values of which
\footnote[1]{The values of the Slater parameters used in this study are
  $F_0 = 7.0$~eV, $F_2 = 12.1$~eV, $F_4 = 7.7$~eV, and $F_6 =
  5.5$~eV. The parameter $F_0$ is often referred to as Coulomb $U$ and
  the combination
\begin{equation*}
\frac{2\ell+1}{2\ell}\sum_{k=1}^\ell F_{2k}
\begin{pmatrix}\ell & 2k & \ell \\ 0 & 0 & 0\end{pmatrix}^2
\end{equation*}
as Hund $J$. With $F_k$ listed above, we have $J=1.0$~eV}
are taken from earlier
investigations~\cite{ogasawara1994,locht2016} where they were adjusted
to reproduce various spectroscopies.
Apart from the limited
accuracy in modeling the strongly correlated 4f electrons, DFT alone also
underestimates the gap between the S~3p and Eu 5d bands. We have corrected
this deficiency by empirically increasing the binding energy of the
S~3p bands such that the final DFT+DMFT band structure is consistent
with valence-band XPS data \cite{cotti1974} as well as with the optical
absorption \cite{guntherodt1971,wachter1972}. Further details of our
DFT+DMFT calculations are reported in \cite{suppl}
(Sec.~\ref{sec:DMFT}) where we also compare our theory with the recent
ARPES measurements revealing changes of the electronic structure
across the ferromagnetic transition \cite{fedorov2021}.

\begin{figure}
\includegraphics[width=\linewidth]{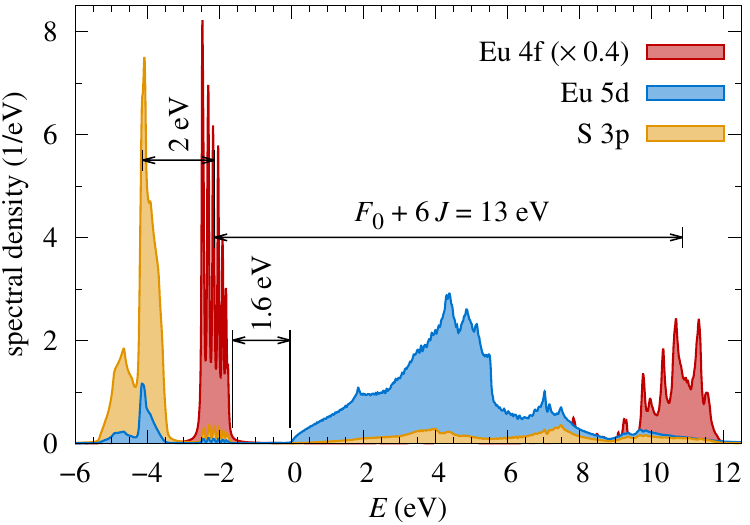}
\caption{\label{fig:DMFT_DOS}Spectral density of the paramagnetic phase
computed with the DFT+DMFT method. Indicated are: the distance between
the S~3p bands and the Eu 4f states (2~eV) as seen in valence-band XPS
\cite{cotti1974}, the optical gap between Eu 4f states and Eu 5d bands
(1.6~eV) as observed in optical absorption \cite{guntherodt1971}, and
the distance between the occupied and unoccupied 4f states (13~eV)
that agrees with the estimate $F_0+6J$ \cite{vandermarel1988}, where
$F_0=7$~eV is the first Slater parameter and $J=1$~eV is the Hund $J$
\cite{Note1}.}
\end{figure}

The orbital-resolved spectral density calculated in the
paramagnetic phase of EuS is shown in Fig.~\ref{fig:DMFT_DOS}. It
reflects the states with one electron removed from the system ($E<0$)
or with one electron added to the system ($E>0$), and it can be
experimentally probed by photoemission and inverse-photoemission
spectroscopy. For uncorrelated states, the spectral density
simplifies to the single-particle density of states. The Eu 4f
states appearing below the Fermi level in Fig.~\ref{fig:DMFT_DOS}
consist of the $L = 3$, $S = 3$ manifold
of the 4f$^6$ configuration, split by the spin-orbit coupling into seven
multiplets ${}^7F_J$ with $J=0,1,\dots,6$. The 4f states
above the Fermi level originate predominantly from the $L = 3$, $S = 3$ manifold
of the 4f$^8$ configuration, the spin-orbit multiplets ${}^7F_J$ are
intermixed in this case by hybridization between the 4f states and
Eu 5d and 6p bands, with which the unoccupied 4f states overlap. The unoccupied
Eu 5d bands are broad and spread over more than 8~eV. This
shape is clearly inconsistent with the main assumption behind the
interpretation of the EuX optical spectra put forward by Wachter \etal
\cite{guntherodt1971}, according to which the Eu 5d$(t_{2g})$ and
5d$(e_{g})$ sub-bands should be sharp and well separated from each other.

The RIXS calculated by means of Eq.~\eqref{eq_RIXS_dip} and using the
occupied and unoccupied 5d DOS plotted in Fig.~\ref{fig:DMFT_DOS}
looks just like the map shown in Fig.~\ref{fig:RIXSmaps}c but with the
features A and B missing (see \cite{suppl},
Fig.~\ref{fig:RIXScomponents}). The feature~C is nicely reproduced in
the calculations. The shape of the computed RIXS follows
entirely from first principles, the energy transfer, at which the
feature~C is
located, comes from the band gaps that were calibrated to valence XPS
\cite{cotti1974} and optical absorption
\cite{guntherodt1971,wachter1972}. This confirms that the RIXS
reported here is compatible with those historical spectroscopic
measurements.

Inspecting the Eu 5d DOS (Fig.~\ref{fig:DMFT_DOS}), it is clear that
the 5d${}\to{}$2p emission involves mainly the Eu 5d states covalently mixed
into the nominally S 3p bands. Hence the feature C can be interpreted
as a charge-transfer feature, corresponding to the final states of the
RIXS process containing a hole in the S~3p valence bands and an electron
excited to the Eu 5d conduction band. There is also a very small
amount of Eu 5d character mixed into the nominally Eu 4f states, which
results in a very faint copy of feature C located at approximately 2~eV
smaller energy transfer -- but its intensity is so small that it is
practically invisible.

%
%

The theory presented so far does not reproduce the experimental
features A and B, which therefore cannot be due to transitions to
extended Eu 5d$(t_{2g})$ and 5d$(e_{g})$ band states as the popular
interpretation of the optical absorption would imply. The sharp nature
of features A and B in 
the experimental RIXS and their lower excitation energy compared
to the S~3p${}\to{}$Eu~5d charge-transfer feature C suggest that they
could be the bound 4f${}^6$5d${}^1$ excitons as hypothesized by Kasuya and
collaborators. To explore this
possibility, we analyzed the basic properties of such excitons with
the aid of the standard DFT combined with the open-core approximation
for the 4f states \cite{malik1977,brooks1991} that allows us to
constrain the 4f shell to a particular filling, 4f${}^6$ or 4f${}^7$
(see \cite{suppl}, Sec.~\ref{sec:opencore} for technical details).

\begin{figure}
\includegraphics[width=\linewidth]{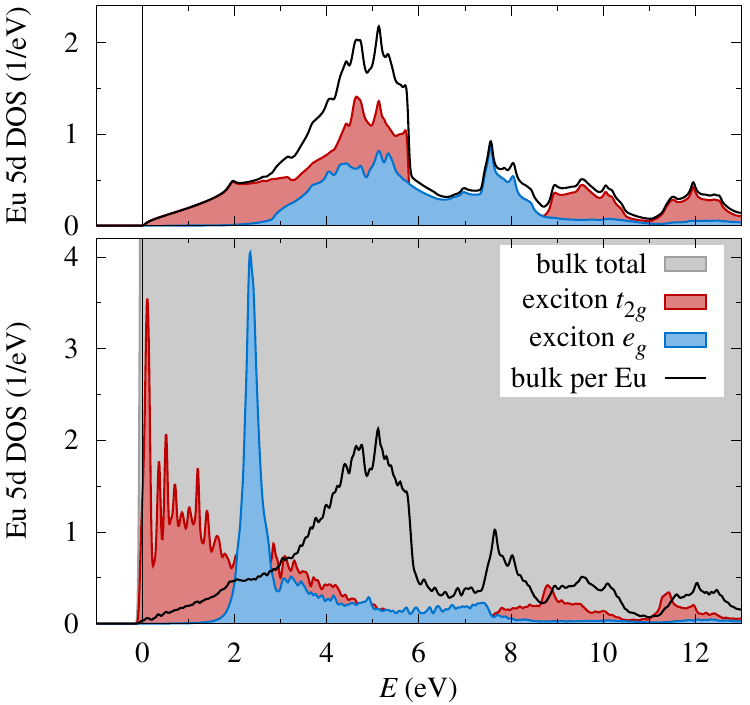}
\caption{\label{fig:exc_DOS}Unoccupied part of the Eu 5d DOS computed in
  the open-core approximation and decomposed into its $e_g$ (blue) and
  $t_{2g}$ (red) components. The bulk
  DOS (top) is compared to the local DOS at the excited atom placed in
  the $4\times4\times4$ supercell (bottom). The grey area corresponds
  to the total 5d DOS in the supercell and indicates that the excitons
  overlap with these bulk states and can decay into them, which is
  needed for compatibility with photoconductivity measurements
  \cite{wachter2013}.}
\end{figure}

The excitons were modeled in supercells (the largest we could afford
was $4\times4\times4$ multiple of the EuS conventional cell, that is,
512 atoms), in which one of the Eu atoms was constrained to have the
4f${}^6$ configuration and the remaining electron was transferred to
the conduction bands. This extra valence electron turns out to remain
bound near the 4f hole, the valence charge density integrated over a
sphere with radius $3.1 r_\text{Bohr}$ around the perturbed atom
contains approximately half of an electron more than the same sphere centered at a
bulk Eu site. Figure~\ref{fig:exc_DOS} then compares the DOS projected
on the 5d states at the atom with the 4f hole to the 5d DOS of the
unperturbed bulk. The open-core bulk DOS is almost identical to the
DFT+DMFT DOS plotted in Fig.~\ref{fig:DMFT_DOS}. The
local 5d DOS at the excited atom substantially deviates from
the bulk 5d DOS: it consists of two distinguished sharp peaks, one of
the $t_{2g}$ and the other of $e_g$ character, each having a weak tail
extending toward high energies. The maxima of these peaks are 2~eV
apart, just like the excitation energies of the observed RIXS
features~A and~B.

To compute the excitation energy $\Delta$ needed to reach the lowest
excitonic state, we evaluate the total energy of the unperturbed
supercell with all atoms in the 4f${}^7$ configuration,
$E_\text{clean}$, and the total energy of the supercell with one of
the atoms constrained to have the 4f${}^6$ configuration,
$E_\text{exc}$. We find
$\Delta=E_\text{exc}-E_\text{clean}=2.4$~eV, that is, very close to
the excitation energy of the RIXS feature A.
Additional aspects of the exciton calculations, in particular the
dependence on the size of the supercell, are discussed in
\cite{suppl}, Sec.~\ref{sec:exciton}.

Finally, we estimate how the computed excitons would show up in
the RIXS plane. To do so, we employ a formula analogous to
Eq.~\eqref{eq_RIXS_dip},
\begin{equation}
\label{eq_RIXS_quad}
I_Q \propto \int \rmd\epsilon\,
\frac{n_{\text{4f}}\,\delta(\epsilon+\Delta)\,
\DOS_{\text{5d exc}}^{\text{empty}}(\epsilon+\omega_1-\omega_2)}%
{(\epsilon-\omega _2-\epsilon_\text{2p})^2+\Gamma_\text{2p}^2/4}\,,
\end{equation}
where the occupied DOS is now the 4f DOS, for simplicity approximated
by a single peak, $n_{\text{4f}}\,\delta(\epsilon+\Delta)$, and the
unoccupied DOS is the local 5d DOS at the atom containing the exciton
(Fig.~\ref{fig:exc_DOS}). Placing the occupied 4f states at the binding energy $-\Delta$
ensures that the lowest exciton appears at the energy transfer equal
$\Delta$ (the unoccupied exciton DOS starts at the Fermi level chosen as the
energy reference, $E_\text{F}=0$). This adjustment of the 4f-state
position can be understood as a many-body correction to the
single-particle (non-interacting) theory, which was used to derive
Eq.~\eqref{eq_RIXS_dip}. In other words, it is a correction due to
the binding energy of the exciton that is by definition zero in the
non-interacting theory.

Using the occupied 4f DOS in place of the occupied DOS in
Eq.~\eqref{eq_RIXS_quad} implies that the emission of the $\omega_2$
photon is due to quadrupole 4f${}\to{}$2p transition, which has a much
smaller intensity than the dipole 5d${}\to{}$2p transition assumed in
Eq.~\eqref{eq_RIXS_dip}. When combining the contributions of
Eqs.~\eqref{eq_RIXS_dip} and~\eqref{eq_RIXS_quad} in
Fig.~\ref{fig:RIXSmaps}c, we assume that the ratio of quadrupolar to
dipolar emission probabilities, $p_Q/p_D=0.024$, is the same as the
ratio of the corresponding absorption probabilities deduced from
absorption data (\cite{bartolome1999} and \cite{suppl},
Sec.~\ref{sec:quadrupolar}). The somewhat counter-intuitive finding of
the dipolar and quadrupolar features having comparable intensity in
the final RIXS map, Fig.~\ref{fig:RIXSmaps}c, stems from the very
different number of the occupied states: there are seven occupied 4f
states available to decay through the quadrupole channel, whereas
there are only about 0.6 occupied 5d states (the integral over the
occupied 5d DOS in Fig.~\ref{fig:DMFT_DOS}) available to decay through
the dipole channel, which results in a large enhancement of the
quadrupole contribution, $n_\text{4f}/n_\text{5d}=11.7$, partly
canceling its small emission probability. It is conceivable that in
more ionic compounds like halides, which have even smaller covalent
admixture of the 5d states in the ligand bands, the quadrupole RIXS
features are even dominant.

The final theoretically derived RIXS map is shown in Fig.~\ref{fig:RIXSmaps}
side by side with the experimental RIXS. The excellent agreement
of the shape and energy position of features A, B and C between
experiment and theory provides a convincing argument for our
interpretation of the RIXS of EuS, and it ultimately demonstrates
that the peaks A and B are the 4f${}^6$5d${}^1$($t_{2g})$ and
4f${}^6$5d${}^1$($e_g)$ localized excitons proposed by Kasuya and
collaborators.

%
%

It is interesting to compare our findings with a recent study by Joos
\etal \cite{Joos2020}. They investigate the Eu$^{2+}$ excited-state
landscape with multiconfigurational $\textit{ab initio}$
embedded-cluster methods and examine the case of Eu$^{2+}$-doped
sulfides MS (M = Ca, Sr, Ba), which have the same rock salt structure as EuS. Indeed,
the optical absorption spectra of the Eu-doped alkaline-earth sulfides
are characterized by peaks similar to A and B of EuS, which were shown
to correspond to the spin-allowed electric-dipole transitions towards
the excited 4f$^6$5d${}^1$($t_{2g}$) and 4f$^6$5d${}^1$($e_g$)
manifolds of Eu$^{2+}$. Both bands posses a complex fine structure
that originates from term and multiplet splitting due to the 4f--5d
Coulomb interaction (that is, the exciton bonding in the terminology of
Kasuya), and the spin-orbit coupling.  This fine structure cannot be
rendered by DFT as it is a single-reference method. Given the
structural and chemical
similarities of EuS and MS:Eu$^{2+}$, it can be presumed that a
similar fine structure is present under features A and B in
Fig.~\ref{fig:RIXSmaps} but it is hidden below the limited
experimental resolution.

In conclusion, we combined RIXS experiments with electronic-structure
calculations to settle a long-standing debate on the nature of low-energy
electronic excitations in the magnetic semiconductor EuS. It was
evidenced that a so-called magnetic exciton is formed in the 1.5~eV to
5.5~eV energy range where the hole and the electron are localized in the
atomic 4f and 5d orbitals of a single Eu$^{2+}$ ion. These excitonic states
correspond to the crystal-field-split 4f$^6$5d$^1$ manifolds that are
known from optical spectroscopy of isolated Eu$^{2+}$ impurities.

\begin{acknowledgments}
Authors acknowledge the ESRF for providing beamtime. Computational
resources were provided by the e-INFRA CZ project (ID:90254),
supported by the Ministry of Education, Youth and Sports of the Czech
Republic. L.A. acknowledges support from the European Research Council
(ERC) under Grant Agreement No.~759696. J.K. acknowledges financial
support by the Czech Science Foundation under the grant No.~21-09766S.
J.J.J. acknowledges the Ghent University Special Research Fund via
project BOF/PDO/2017/002101.
\end{acknowledgments}

\bibliography{EuS_vtc_RIXS,suppl}

\end{document}